\documentclass[a4paper,11pt]{article}
\pdfoutput=1 

\usepackage{jinstpub} 

\usepackage{lineno}

\title{Timing techniques with picosecond-order accuracy for novel gaseous detectors }

\author[1]{A. Tsiamis,\note{Corresponding author.}}
\author{K. Kordas,}
\author[2]{I. Manthos,\note{Now at University of Birmingham.}}
\author{M. Tsopoulou}
\author{and S.E. Tzamarias}

\affiliation{Department of Physics, Aristotle University of Thessaloniki,\\University Campus,GR-54124,Thessaloniki, Greece}

\emailAdd{angelos.tsiamis@cern.ch}

\abstract{A simulation model is developed to train Artificial Neural Networks (ANN), for precise timing of PICOSEC Micromegas detector signals. The aim is to develop fast, online timing algorithms as well as minimising the information to be saved during data acquisition. PICOSEC waveforms were collected and digitised by a fast oscilloscope during a femptosecond-laser test beam run. A data set comprising waveforms collected with attenuated laser beam intensity, eradicating the emission of more than one photoelectron per light pulse from the PICOSEC photocathode, was utilised by a simulation algorithm to generate waveforms to train an ANN. A second data set of multi-photoelectron waveforms was used to evaluate the ANN performance in determining the PICOSEC Signal Arrival Time, relative to a fast photodiode time-reference. The ANN timing performance is the same as the results of a full offline signal processing, achieving a timing precision of 18.3$\pm$0.6\,ps.}


\proceeding{12$^{\text{th}}$ International Conference on Position Sensitive Detectors \\
 12 - 17 September 2021.\\
University of Birmingham, UK}

\notoc

\begin{document}
\maketitle
\flushbottom

\section{Introduction}
\label{sec:intro}
\begin{figure}[h] 
\centering
\begin{minipage}{.4\textwidth}
 \includegraphics[width=.9\linewidth]{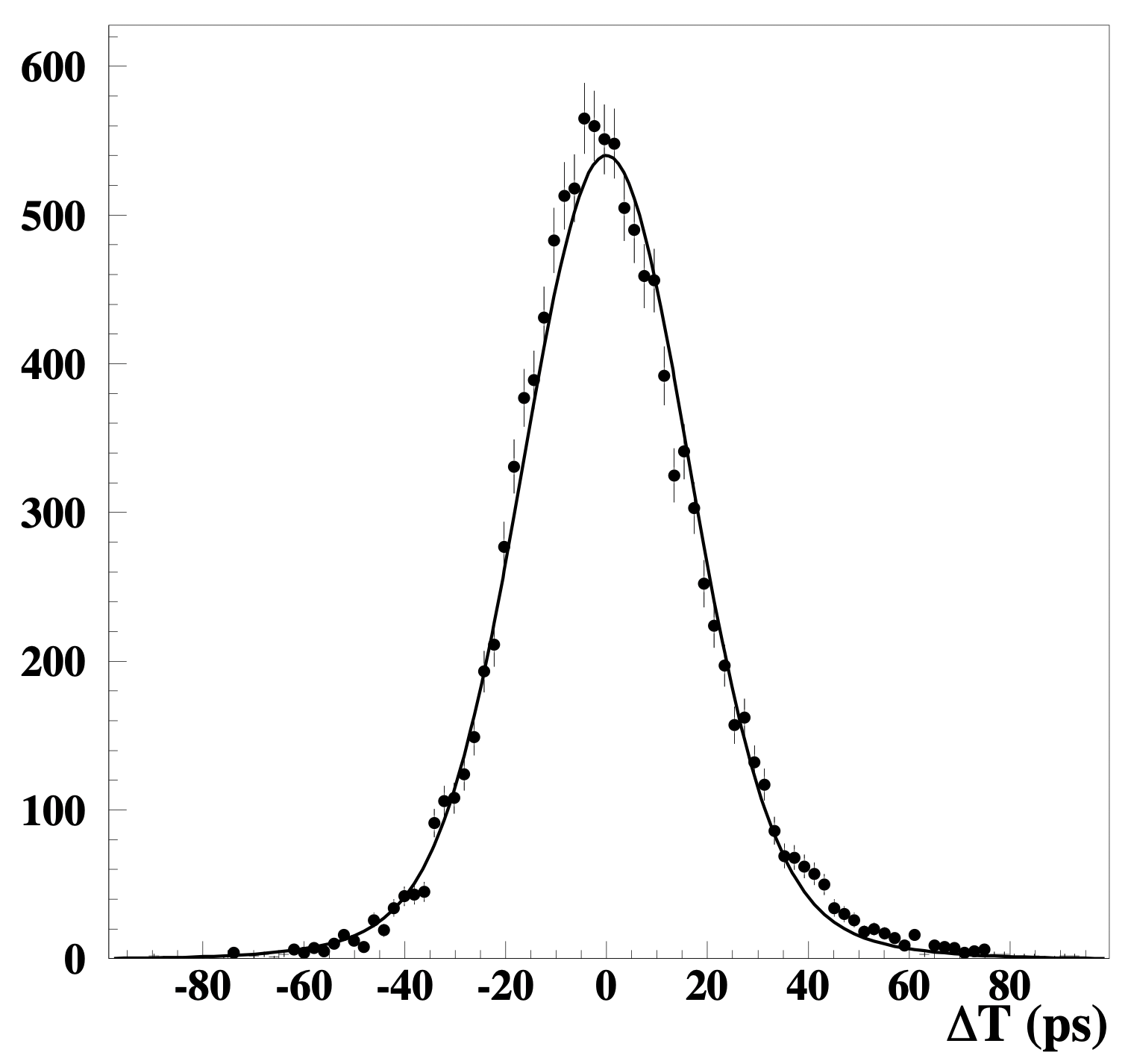}
\end{minipage}
\centering
\begin{minipage}{.4\textwidth}
\includegraphics[width=.9\linewidth]{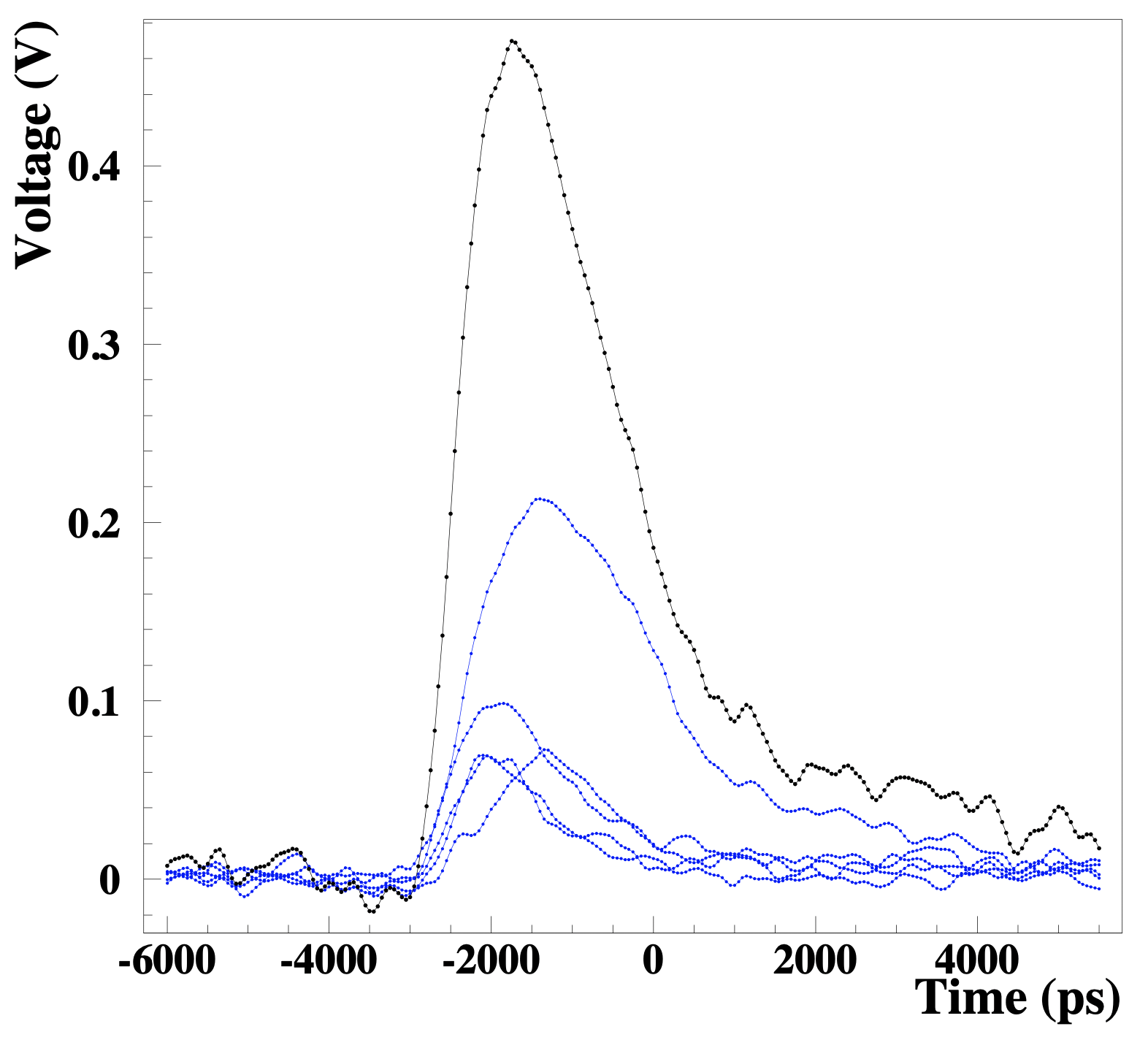}
\end{minipage}
\caption{(Left) The distribution of the PICOSEC SAT - evaluated in the offline analysis - of all available pulses. The RMS of this distribution determines the global timing resolution at 18.3$\pm$0.6\,ps. (Right) Simulation of the PICOSEC response to  5\,pes by summing SPE-set waveforms. The single-pe waveforms have been synchronised using the respective time reference signals provided by the photodiode and a 3$^{\textrm{rd}}$ degree polynomial interpolation between digitizations.  \label{fig:fig1}}
\end{figure}

The PICOSEC Micromegas detector has the potential for precise timing at the picosecond level \cite{a,b}. In this work we develop a simulation model to train Artificial Neural Networks (ANN) for precise timing of PICOSEC \cite{c} signals. The aim is a fast online signal processing, as well as the minimisation of the necessary information to be saved during data acquisition. Two sets of PICOSEC waveforms were collected and digitized by a fast oscilloscope (20\,GS/s) during a femtosecond-laser test beam run, with the PICOSEC operating at stable conditions \citep{b}. In both sets a fast photodiode was used to time the laser pulses, with <\,3\,ps resolution, and to provide an accurate time-reference for each PICOSEC digitized waveform. The first data set (SPE-set) comprises waveforms collected with attenuated laser beam intensity, eliminating the production of more than one photoelectrons (pes) on the PICOSEC photocathode. Section \ref{sec2} describes the utilisation of the SPE-set in generating simulated waveforms for training the ANN. The second data set (EXP-set) comprises waveforms corresponding to many photoelectrons produced by the same laser pulse. The  number of pes per laser pulse, found to follow a Poisson distribution with a mean value of $\approx7.8$. In Section \ref{sec3}, the EXP-set was used to evaluate the ANN performance in determining the PICOSEC Signal Arrival Time (SAT), relative to the photodiode time-reference. The ANN timing precision is compared to the results of a full offline timing analysis \cite{d}, i.e. the application of the Constant Fraction Discrimination (CFD) method to a logistic function that fits the leading edge of the digitised Electron Peak. We demonstrate that the ANN timing achieves the same precision as the one provided by the full offline analysis (i.e. 18.3$\pm$0.6\,ps, Figure \ref{fig:fig1} - left).

\section{Simulation of multi-pe pulses using single-pe waveforms}\label{sec2}

\begin{figure}[h] 
\centering
\begin{minipage}{.4\textwidth}
\includegraphics[width=.9\linewidth]{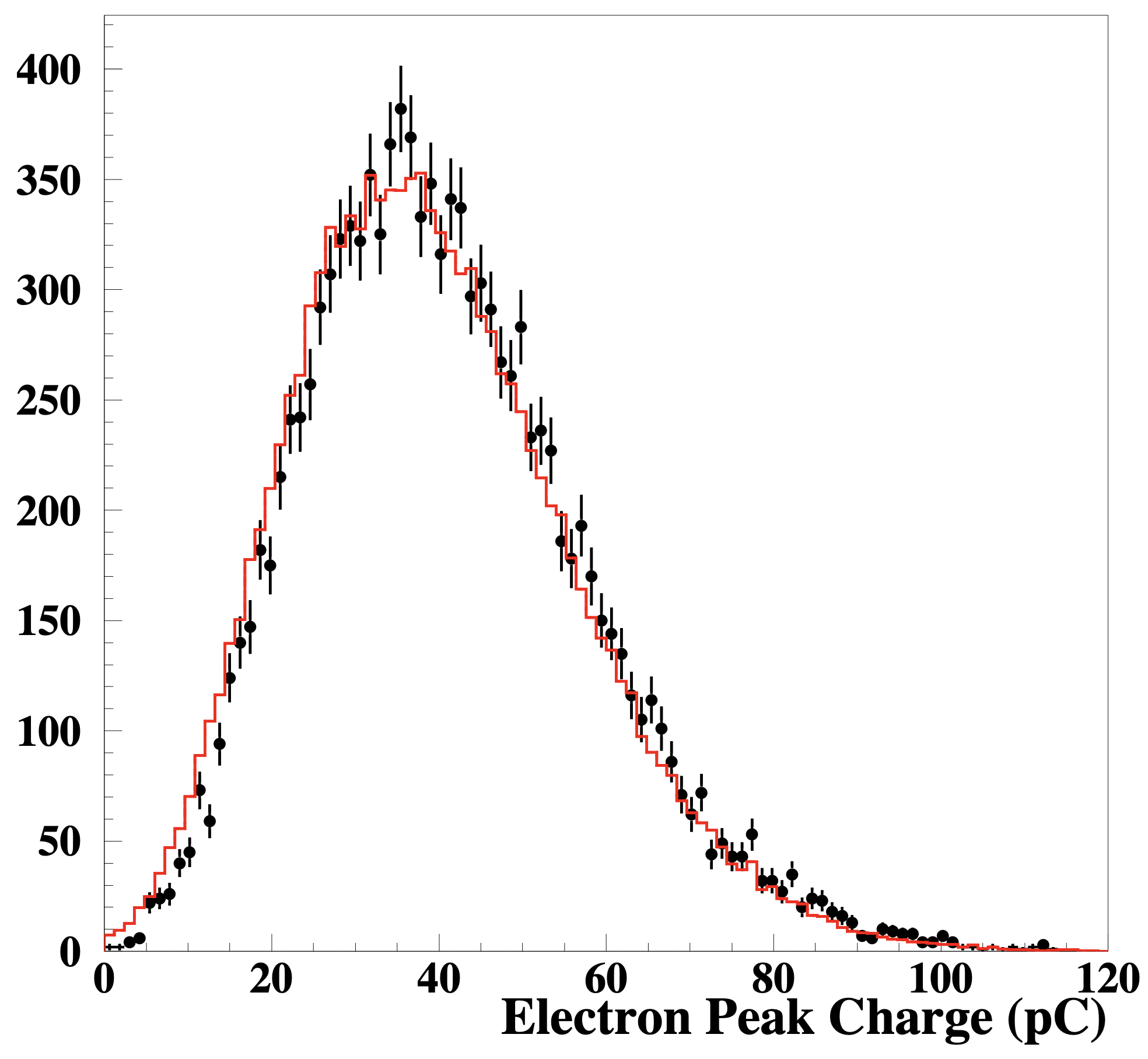}
\end{minipage}
\centering
\begin{minipage}{.4\textwidth}
\includegraphics[width=.9\linewidth]{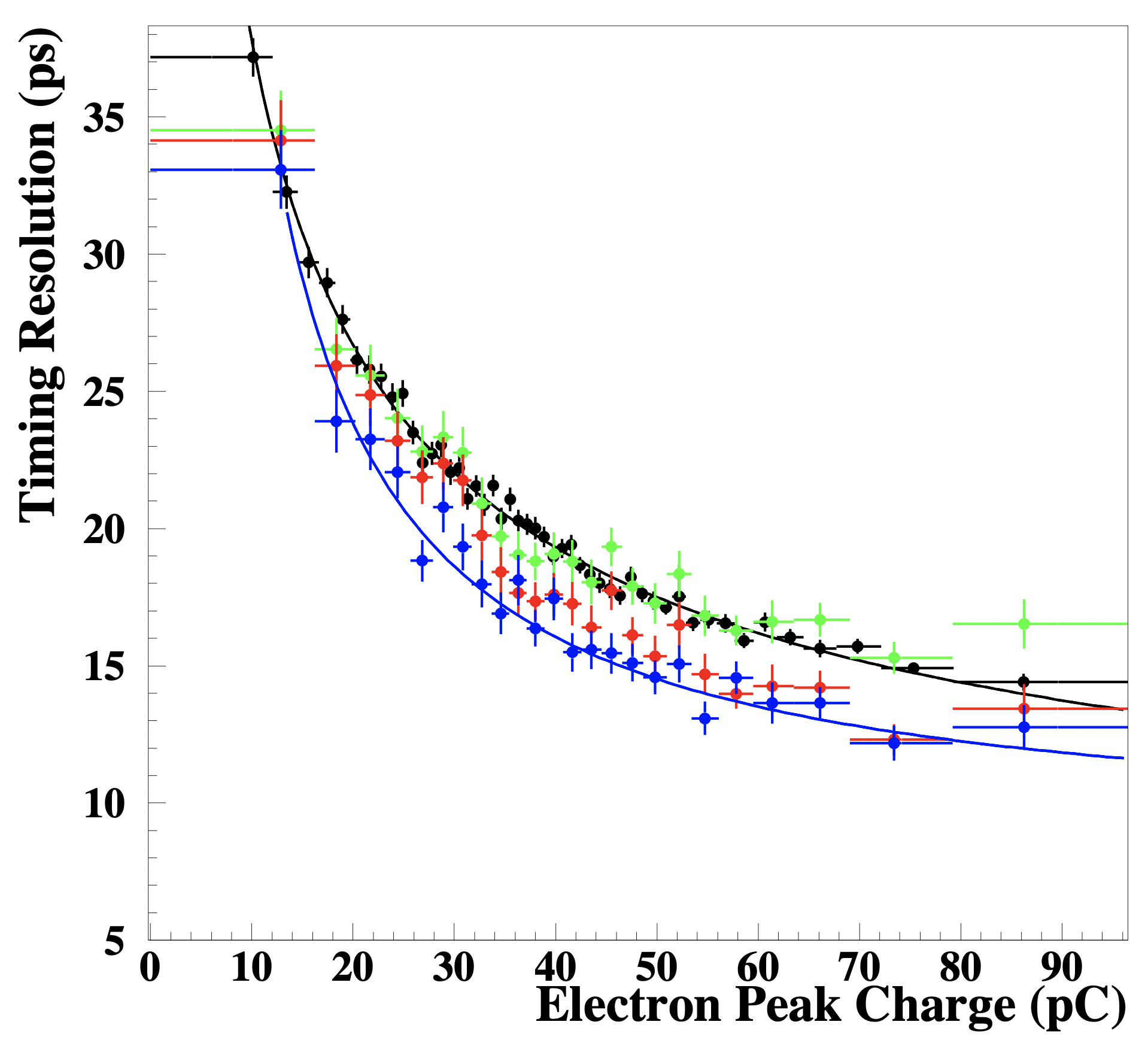}
\end{minipage}
\caption{(Left) The charge distribution of the EXP-set Electron Peaks (black points) and the simulation prediction (red histogram). (Right) The timing resolution as a function of the Electron Peak charge of: (black) the simulated pulses, (blue) the EXP-set waveforms, (red) after adding appropriately noise to the EXP-set waveforms, and (green) after adding the timing reference resolution (2.8\,ps per single-pe pulse).\label{fig:fig2}}
\end{figure}

The waveforms of the SPE-set are used to simulate signals corresponding to the PICOSEC response to many pes. Figure \ref{fig:fig1} (right) demonstrates the emulation of the PICOSEC response to 5 pes by adding single pe waveforms. The deconvolution algorithm, described in \citep{e}, is used 
to estimate the number of pes that produce the EXP-set waveforms. The number of pes (N) found to follow a Poisson distribution, with mean value of 7.8$\pm$0.1\,pes. 
The EXP-set was then simulated by summing up N single-pe waveforms, chosen randomly, where N follows the aforementioned Poisson distribution. The simulated pulses were then analysed as the real data. The charge distribution of the simulated pulses agrees very well with the respective EXP-set distribution, as shown in Figure \ref{fig:fig2} (left).
Figure \ref{fig:fig2} (right) presents results of the full timing analysis of the simulated pulses. The global timing resolution found to be 21.3$\pm$0.6\,ps, which is by 3\,ps worse than the precision found in the real data. This small difference is due to the fact that by summing N single-pe pulses a) we accumulate extra noise on the simulated waveforms, and b) we add (quadratically) N times the time jitter of the photodiode signal. 

\section{Timing with Artificial Neural Networks}\label{sec3}

\begin{figure}[h] 
\centering
\begin{minipage}{.4\textwidth}
\includegraphics[width=.9\linewidth]{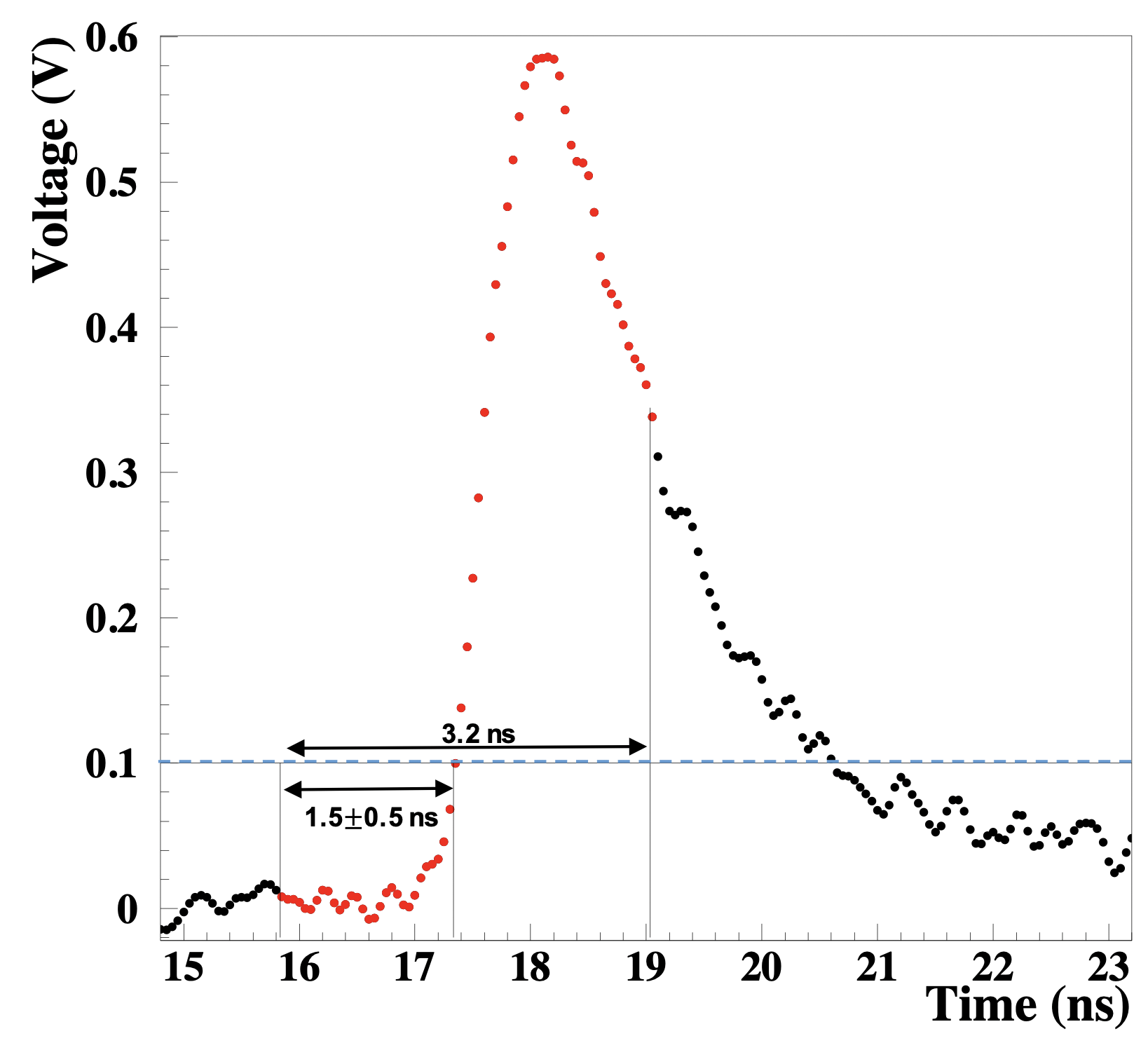}
\end{minipage}
  \centering
\begin{minipage}{.4\textwidth}
 \includegraphics[width=.9\linewidth]{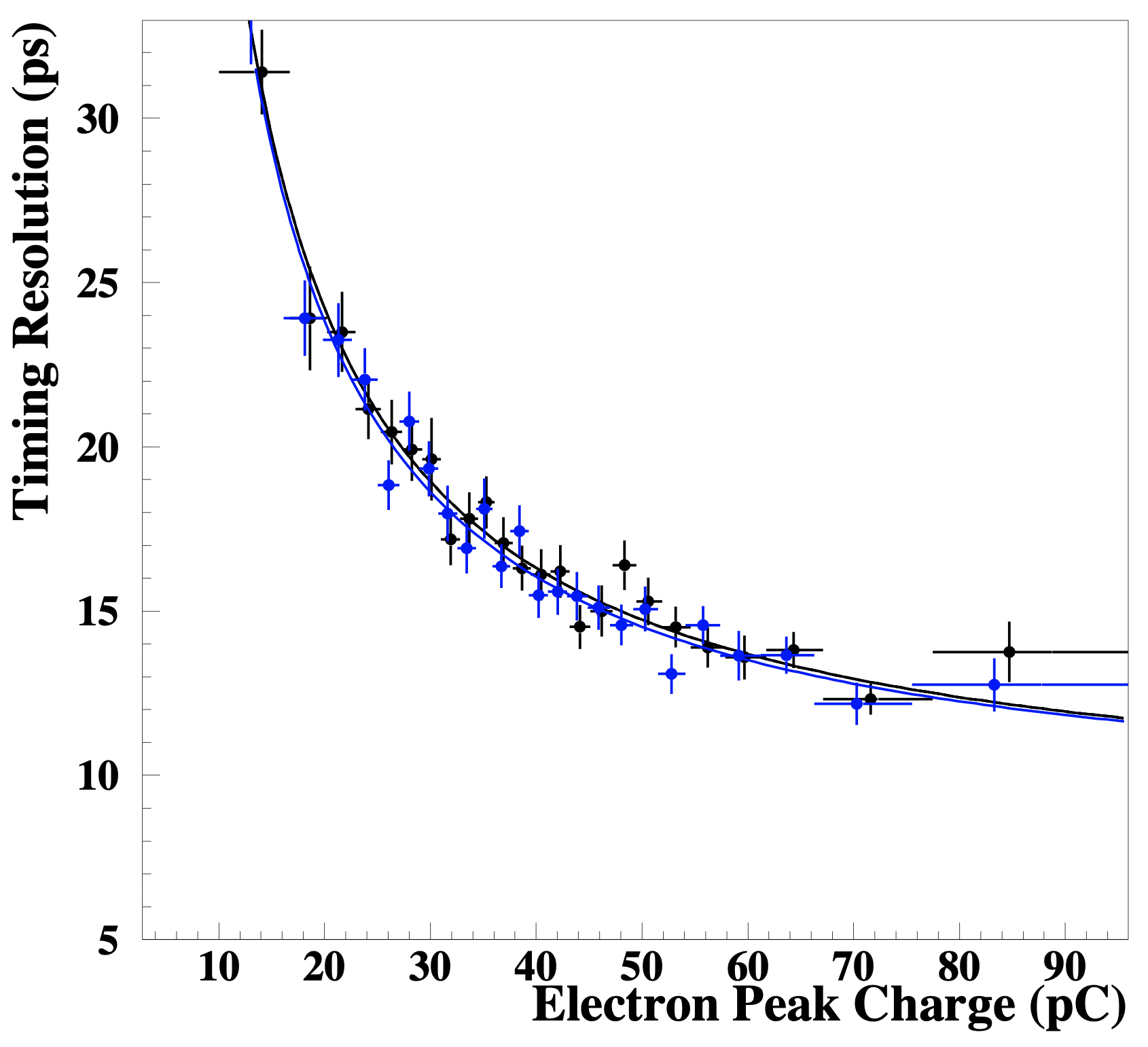}
\end{minipage}
  \caption{(Left) A typical PICOSEC digitised waveform. The red points denote the digitised information that is presented to the ANN. In this example, the ANN is fed with 64 inputs.
(Right) The timing resolution using ANN (black) and full analysis of the Electron Peak waveforms (blue), as a function of the Electron Peak charge. A 18.3$\pm$0.6\,ps timing resolution is achieved in both cases. \label{fig:fig3}}
\end{figure}

A feedforward ANN has been employed to extract the timing information from the leading edge of the PICOSEC waveform Electron Peak. It is assumed that a threshold trigger, at 100\,mV, is used to select waveforms and to provide a timestamp with a time jitter of $\pm$500\,ps. This timestamp is used to select digitizations in a time window of 3.2\,ns as shown in Figure \ref{fig:fig3} (left). During the training session, the digitizations of the simulated Electron Peak waveforms are given as inputs to the ANN together with their respective reference (photodiode) times as the respective target values. After learning, the waveforms of the EXP-set are presented as inputs to the network, and its output is compared with the respective time-references to evaluate the ANN performance. As shown in Figure \ref{fig:fig3} (right), the ANN timing achieves the same accuracy as the full signal processing analysis (18\,ps). When the ANN is fed with 16 inputs (i.e. corresponding to signal digitisation with 5\,GS/s sampling rate) achieves 19.2$\pm$0.8\,ps timing resolution, in agreement with the results of the offline signal processing. Extensive tests have shown that the ANN timing performance is unbiased, e.g. is not sensitive to the selection of the first digitization, does not depend on the charge distribution of the Electron Peaks, etc. 

\section{Conclusions}
A model has been developed to simulate the PICOSEC response to many pes by using single-pe waveforms that have been collected and digitized in a laser test beam run. The simulation model is used to generate learning samples for an ANN. Although the simulated waveforms are not completely mutually-uncorrelated, suffer extra noise and a small systematic timing error, they contain the necessary information that the ANN needs to learn and perform precise timing. Extensive tests, reported in a forthcoming publication, demonstrated the ability of the ANN to time PICOSEC waveforms with the same precision as the one achieved with a full signal processing analysis.

\acknowledgments

The authors would like to thank kindly the CERN-RD51 PICOSEC Micromegas collaboration for providing experimental data from the PICOSEC detector.


\end{document}